\documentclass[final,numberedheadings]{aipproc}
\usepackage{epsfig,amsmath,amssymb,mathrsfs,bm}
\usepackage[active]{srcltx}
\layoutstyle{6x9} 
\def\figdir{./FQXicontest2012/}
\renewcommand\leq{\leqslant}
\def\vec#1{{\mathbf{#1}}}
\begin{document}
\title[]{The Dirac Quantum Automaton: a preview\footnote{Work presented at the conference {\em
      Quantum Theory: Reconsideration of Foundations-6} on June 12th 2012 at Linnaeus
    University, V\"axj\"o, Sweden.}}\classification{03.65.-w}\keywords {Foundations of Physics,
  Axiomatics of Quantum Theory, Special Relativity, Quantum Field Theory} \author{Giacomo Mauro
  D'Ariano}{address={{\em QUIT} Group, Dipartimento di Fisica, Universit\`a di Pavia, via Bassi 6,
    27100 Pavia,
    Italy, {\em http://www.qubit.it}\\
    Istituto Nazionale di Fisica Nucleare, Gruppo IV, Sezione di Pavia}}
\begin{abstract} Quantum Information and the new informational paradigm are entering the domain of
  quantum field theory and gravity, suggesting the quantum automata framework. The quantum automaton
  is the minimal-assumption extension to the Planck and ultrarelativistic scales of quantum field
  theory. It can describe localized states and measurements, which are unmanageable by quantum field
  theory. The automaton theory is a very promising framework for quantum gravity, since it is
  quantum {\em ab-initio}, with relativistic covariance as emergent and not assumed a priori, it is
  free from all the problems arising from the continuum, it doesn't suffer violations of causality,
  and has no divergences. It is the natural scenario to accommodate the quantum holographic
  principle.  Lorentz covariance and all other symmetries are violated in the ultrarelativistic
  Planckian regime, but are perfectly recovered at the Fermi-scale.

  In the present report, after briefly reviewing the fundamental principles at the basis of the
  quantum cellular automata extension of quantum field theory, I will present a preview of recent
  results on the Fermi scale limit \cite{BDT} and on the Dirac automaton in two space-dimensions
  \cite{MP}.  The automaton in three dimensions is under way.
\end{abstract}
\maketitle
%%%%%%%%%%%%%%%%%%%%%%%%%%%%%%%%%%%%%%%%%%%%
%% MAINMATTER
%%%%%%%%%%%%%%%%%%%%%%%%%%%%%%%%%%%%%%%%%%%%

\section{Introduction}

After the recent success of deriving the full structure of quantum theory  \cite{our, viewpoint,
  hardy,masanes} from information-theoretical principles, the informational program for physics now
continues by entering relativistic quantum field theory (QFT), with the main motivation of having a
complete framework that can harbor a quantum theory of gravity. The program leads to a theory that
is {\em quantum ab-initio}, made of discrete quantum systems in interactions, with the dynamics
described by a quantum cellular automaton (QCA), namely a plaquette of locally connected gates that
translationally reproduce an infinite quantum circuit. The plaquette embodies the physical law, and
takes the place of the Lagrangian density in field theory.

Why should we take the QCA as a promising framework for quantum gravity? For a long list of reasons:
{\em (a)} The QCA relaxes the tension between quantum theory and relativity, since it does not need
the latter, which is emergent from the former; {\em (b)} The QCA does not need quantization rules,
being quantum {\em ab-initio}, with the automaton derived from first principles and from the
symmetries of the interaction topology; {\em (c)} The QCA is free from all the problems that plague
QFT arising from the continuum, a tremendous theoretical advantage compared to QFT;  {\em (d)} The
QCA doesn't suffer violations of causality, and has no superluminal tails of the wavefunction;  {\em
  (e)} The QCA is the natural scenario to accommodate the quantum holographic principle---the
mysterious informational principle that seems to be at the basis of a ``thermodynamical'' theory of
gravity \cite{jacobson,verlinde}.

In the preliminary studies \cite{Vaxjo2010,pla,Vaxjo2011} we have seen some of the power of the QCA
approach. Despite its simplicity, the QCA is pregnant with physical content, leading to unexpected
interesting predictions, e.g. the Dirac automaton--the most elementary theory of this
kind--anticipates a maximum mass for the Dirac particle \cite{Vaxjo2010,pla}, just as a consequence
of the unitariety of quantum evolution, without invoking black-hole arguments from
general-relativity. It also opens completely new routes for redefining the crucial mechanical notion
of inertial mass through a natural cinematical definition, also providing a redefinition of the
Planck constant (see also Ref.  \cite{Saggiatore} for a didactical review).

The first preliminary ideas about the QCA extension of QFT were introduced at Perimeter Institute on
Feb. 2010  \cite{Pirsa}, and then at the V\"axj\"o in June  \cite{Vaxjo2010}. Last year in V\"axj\"o I
presented the quantum automaton of the Dirac field  \cite{Vaxjo2011} in one space dimensions, and 
showed how the dynamics of quantum particles emerges from the quantum processing of the automaton,
which reproduces the Dirac field dynamics at the Fermi scale, while departing from it at the
Planck-scale.  

In which sense the quantum automaton represents an extension of QFT? Because it describes also
localized states and measurements that are not representable within QFT.  Relativistic covariance
and other symmetries are violated, but are perfectly recovered in the thermodynamic limit
corresponding to the Fermi-scale. 

What are the information-theoretical principles that replace relativity in the QCA? As I will
comment in this paper, in addition to the causality principle (the first of the six axioms of
Quantum Theory of Ref.  \cite{our}), we have: 1) the Quantum Church-Turing principle; 2) the
principle of topological homogeneity of interactions. These principles have been introduced in Ref.
 \cite{Saggiatore} and discussed versus the relativity principle in my FQXi essay of this year
 \cite{myFQXi12}.

 In the present nontechnical report, in addition to briefly reviewing the principles and the main
 features of the QCA extension of QFT, I will give some snapshots of recent results on the Dirac
 automaton in two space-dimensions \cite{MP}.  Two long technical manuscripts containing the
 analytical derivation of the asymptotics at the Fermi scale of the automaton \cite{BDT} and the
 automaton for three space dimensions \cite{MP} are under way, in collaboration with Paolo
 Perinotti, Alessandro Bisio and Alessandro Tosini: these will be the topics of next conference in
 V\"axj\"o in 2013.

\section{The substitutes for the relativity principle}
The Quantum Theory derived from purely informational principles in Ref.  \cite{our} is just the {\em
  abstract theory of systems}, including the mathematical framework of Hilbert spaces, the algebra
of observables, and the unitary transformations, but it has no bearing on the {\em mechanics} of
particles and quantization rules, which we include in the so-called ``Quantum Mechanics''.  The
latter, however, is just a small portion of the general quantum field theory, which is again a
theory of systems (the quantum field modes), and the only remaining mechanical elements are the
field quantization rules, or equivalently, the path-integral. In order to have a theory autonomous
from classical mechanics, we need to avoid the quantization rules, whereas, on the contrary, we want
to recover classical mechanics as emergent, and have a corresponding {\em classicalization rule}.
But, how to formulate a theory that is quantum ab initio?  We need new principles in addition to the
six ones of quantum theory. Those principles, which will be the substitute of the relativity
principle, are themselves of informational nature. They are: 1) the Quantum Church-Turing principle;
2) the principle of topological homogeneity.

\subsection{The Quantum Church-Turing principle} 
Rephrasing D. Deutsch  \cite{Deutsch}: ``Every physical process describable in finite terms must be
perfectly simulated by a quantum computer made with a finite number of qubits and a finite number of
gates''. In an operational framework based on a specularity between experimental protocols and
theoretical algorithms, we say more precisely:

\begin{itemize}
\item[] {\bf\em The Quantum Church-Turing principle:} {\em Every finite experimental protocol is perfectly
    simulated by a finite quantum algorithm}.
\end{itemize}

It is immediate to see that the principle implies two
sub-principles: 
\begin{itemize}
\item[a)]  {\em the density of information is finite}, 
\item[b)] {\em interactions are local}.
\end{itemize}
The kind of information that we are considering here is quantum, whence the assertion that the
density of information is finite means that the dimension of the Hilbert space is finite.  This
means that e.g.  the {\em boson} should be regarded as an asymptotic emergent notion.  The finite
dimension of the Hilbert space also implies locality of interactions, namely that the number of
quantum systems connected to each gate is finite.

Richard Feynman is reported to like the idea of finite information density. He felt that ``There
might be something wrong with the old concept of continuous functions. How could there possibly be
an infinite amount of information in any finite volume?'' \cite{book}.

\subsection{Principle of topological homogeneity of interactions}  
\begin{itemize}
\item[] {\bf\em Principle of topological homogeneity of interactions:} {\em the quantum algorithm describing
    a physical law is a periodic quantum network}.
\end{itemize}
In the informational paradigm the physical law is represented by a finite set of connected quantum
gates, corresponding to a finite protocol, theoretically specular of a finite quantum algorithm.
Thus locality of interactions is required in order to define a physical law in terms of a finite
protocol under the local control of the experimenter, whereas homogeneity represents the
universality of the law, which is assumed to hold everywhere and ever.  It follows that
algorithmically the physical law is represented by the quantum unitary cellular automaton of Werner
and Schumacher \cite{Werner}. The ``space''-period and the ``time''-period of the automaton
correspond to the minimum space and time units $l_P$ and $t_P$--the Planck distance and the Planck
time, respectively. At some very small scale--the Planck scale--the world is discrete!

\section{The quantum cellular automaton}
Causality together with the Quantum Church-Turing principle imply that information propagates at
finite speed, the maximum speed being the speed of light $c=l_P/t_P$, corresponding to the causal
speed of the automaton. The two principles together imply that the state of any finite set of
systems can be evaluated exactly as the evolution for finitely many time-steps of a larger, yet
finite, number of systems in the past causal cone, regardless the quantum network being unbounded.
Hence, as long as we evaluate the evolution for finite number of steps, everything is exact and
finite, and we don't need boundary conditions. We take as {\em vacuum state} any state that is
locally invariant under the automaton evolution. The {\em localized states} are those that differ
from the vacuum only for a finite number of systems.  The future causal cone of such
state-supporting systems is then the only place where we need to evaluate the evolution.  Evidently
we get no divergence, nor ultraviolet nor infrared, since there is no continuum, and the evaluation
runs for finitely many systems: the Quantum Church-Turing principle excludes {\em tout court} the
continuum and the infinite dimension.

\subsection{Recovering quantum field theory}
The old field theory is recovered as an approximation via an analytical asymptotic evaluation of the
automaton evolution in the Fermi-scale thermodynamic limit of infinitely many steps and delocalized
states, the latter corresponding to customary quantum particles. The momentum is given by taking the
superposition of localized states in neighboring positions with a fixed relative phase-shift. In
this way we obtain back the usual Dirac equation in the relativistic regime of small masses and
momenta, but at the same time we also describe the physics of very large Planckian masses and
ultrarelativistic huge momenta.  

Recently an analytical method to derive the asymptotic dynamics for large number of steps and
smoothly varying states (of the Schwartz class, as those of QFT) has been derived \cite{BDT}.
Besides re-obtaining the Dirac equation for small inertial mass and small momenta, we can describe
analytically the automaton dynamics very precisely for Schwartz states with narrow band around
momentum $k=k_0$ in a large range of momentum and mass by a $k$-dependent Schr\H{o}dinger equation
with drift term. For example, corresponding to the unitary matrix of the Dirac automaton for $d=1$
\cite{Vaxjo2011}
\begin{equation}
\mathbf{U}=\begin{pmatrix}
n S & im \\ im & nS^\dag \end{pmatrix},
\end{equation}
with $0\leq m\leq 1$ the inertial mass, $n=\sqrt{1-m^2}$, and $S\bm{\psi}(x)=\bm{\psi}(x+1)$
shifting the automaton state to the left, one has the Schr\H{o}dinger equation \cite{BDT}
\begin{equation}
i\partial_t \tilde{\phi}(x,t)= s\left(i v \frac{\partial}{\partial x}
    -\frac{1}{2}D \frac{\partial ^2}{\partial x^2} \right)\tilde{\phi}(x,t),
\end{equation}
where $\tilde{\phi}(x,t)=\phi(x,t)e^{-i[k_0 x-\omega(k_0)t]}$, $\omega(k)$ the dispersion relation
of the automaton, $\phi(x,t)$ the amplitude of the superposition, $s=\pm 1$,
$v(k_0,m)=\sqrt{\frac{n}{1+m^2\cot^2(k_0)}}$, and
$D(k_0,m)=\frac{nm^2\cos{(k_0)}}{(\sin^2(k_0)+m^2\cos^2(k_0))^{\frac32}}$.

\subsection{Space-time metric emerging from topology} 
The homogeneity of interactions is a topological property, not a metrical one: the interactions make
a network of connected systems where the length of each link of the graph has no physical meaning.
Space-time metric emerges from the pure topology by event counting: the Planck length $l_P$ and time
$t_P$ conceptually are the just digital-analog conversion factors. In this fully Pythagorean digital
world, also the particle mass $m$ of the Dirac automaton is a pure number $0\leq m\leq 1$, and the
Planck mass $m_P$ represents the conversion factor to the customary kilogram unit.

\subsection{Universal constants of QCA theory}
The adimensional mass is the only parameter characterizing the Dirac automaton, whereas the three
quantities $l_P,t_P,m_P$ are the irreducible universal constants for the QCA theory. Such system of
universal constants directly defines $[L]$, $[T]$, and $[M]$ units. Moreover, no square root is
needed to express the other universal constants in terms of $l_P,t_P,m_P$, namely the system of
QCA universal constants is {\em fundamental} in the Wilczek's sense \cite{Wilczek}. Very interesting are
the relations giving the Planck constant and the Gravitational constant as derived quantities:
\begin{equation}
\hbar=m_Pc^2t_P, \quad\text{(Planck constant)},\qquad
G=l_Pc^2/m_P, \quad\text{(Gravitational constant)}.
\end{equation}

\begin{figure}[ht]
\includegraphics[width=.6\textwidth]{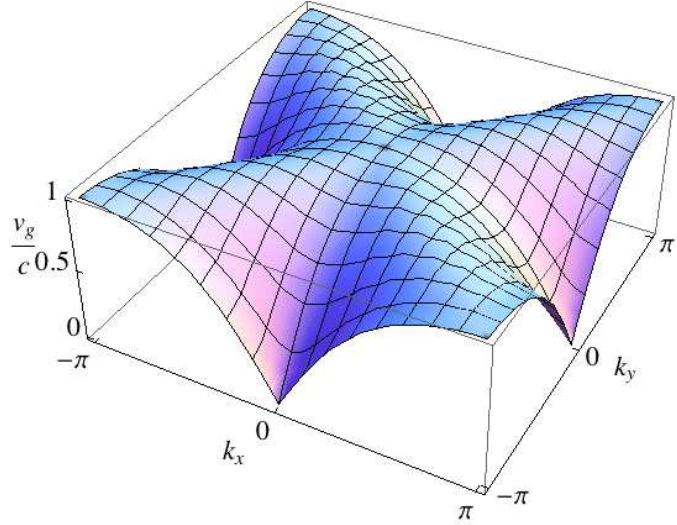}
  \caption{Group velocity $v_g$ (normalized to $c$) for a zero-mass particle automaton
    versus the adimensional momentum $(k_x,k_y)$ (from Ref.   \cite{MP}). The speed is approximately
    isotropic for low momentum (relativistic regime), and becomes anisotropic for very large momenta
    (ultrarelativistic regime). (Reprinted from Ref.  \cite{myFQXi12}).}\label{f:gv}
\end{figure}

\subsection{Inertial mass}
As I already explained in my 2011 FQXi essay \cite{myFQXi11}, in the one-dimensional Dirac quantum
automaton the inertial mass is reinterpreted as the slowing down factor of the information flow via
the coupling between the two modes flowing along the network directions of maximal speed $c$. For
space-dimensions $d>1$ the inertial mass retains its cinematical definition, being the coupling
between the two different chiralities at maximal speed  \cite{MP}. For $d=2$ the unitary matrix of
the automaton is given by  \cite{MP}
\begin{equation}\label{auto2d}
\mathbf{U}=\begin{pmatrix}
n\begin{pmatrix} S_{xy}^+ &\omega^* S_{xy}^-{}^\dag\\
-\omega S_{xy}^-& S_{xy}^+{}^\dag\end{pmatrix}
& im I\\ imI &
n\begin{pmatrix} S_{xy}^+{}^\dag &-\omega^* S_{xy}^-{}^\dag\\
\omega S_{xy}^-& S^+_{xy}\end{pmatrix}\end{pmatrix},
\end{equation}
where $|\omega|=1$, and $S_{xy}=S_x\pm S_y^\dag$, with $S_\alpha$ shifting the automaton state in
the $\alpha$ direction on the topological interaction-network.

\subsection{Particle speed and Planck mass as bound on mass} 
The dispersion relation of the automaton in Eq. (\ref{auto2d}) is given by  \cite{MP}
\begin{equation}
\omega(k)=\pm\arcsin[\tfrac{1}{2}n(\cos k_x+\cos k_y)].
\end{equation}
In Fig. \ref{f:gv} the corresponding group velocity for a zero-mass particle automaton is reported
versus the adimensional momentum $k$ for $m=0$ (i.~e. $n=1$). One can see that the speed is
approximately isotropic for low momentum in the relativistic regime, whereas it becomes anisotropic
for very large momenta in the ultrarelativistic regime, approaching zero at the Planckian
wavelengths ${\bf k}=\pm(\pi, 0)$ or ${\bf k}=\pm(0,\pi)$ at the Brillouin zone border. For massive
particles the speed of light in the Dirac equation decreases also versus the mass for very large
Planckian masses, the automaton evolution becoming stationary at $m=1$, i.~e. at the Planck mass
\cite{pla}, since for larger masses the evolution would be non unitary. The Planck mass is thus the
maximum possible mass for a Dirac particle. As already mentioned, the existence of a bound for the
mass of the Dirac particle does not follows from general relativistic arguments (occurrence of a
mini black holes), but just from quantum theory.

\begin{figure}[ht]
\begin{tabular}{cc}
\includegraphics[width=.46\textwidth]{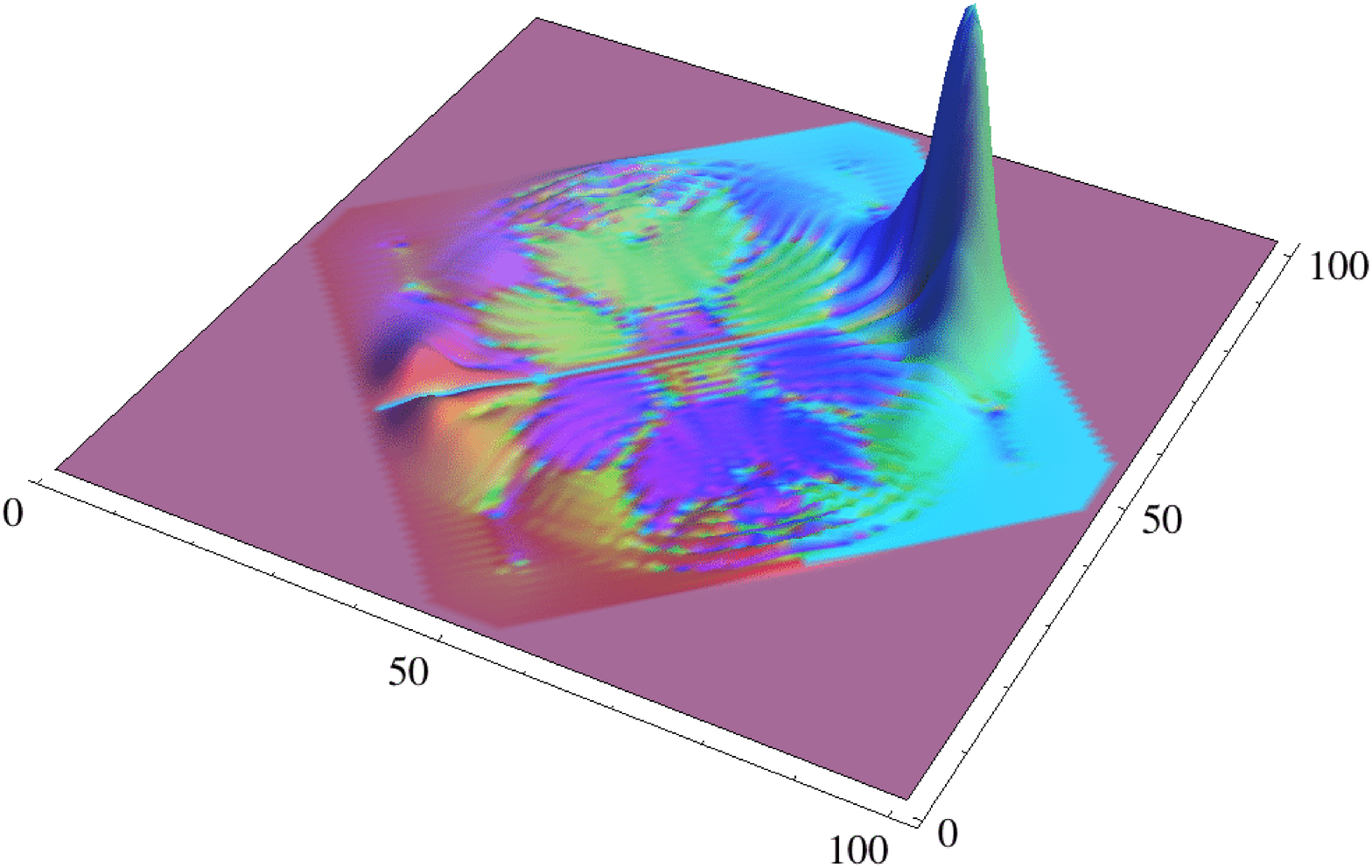}&\includegraphics[width=.46\textwidth]{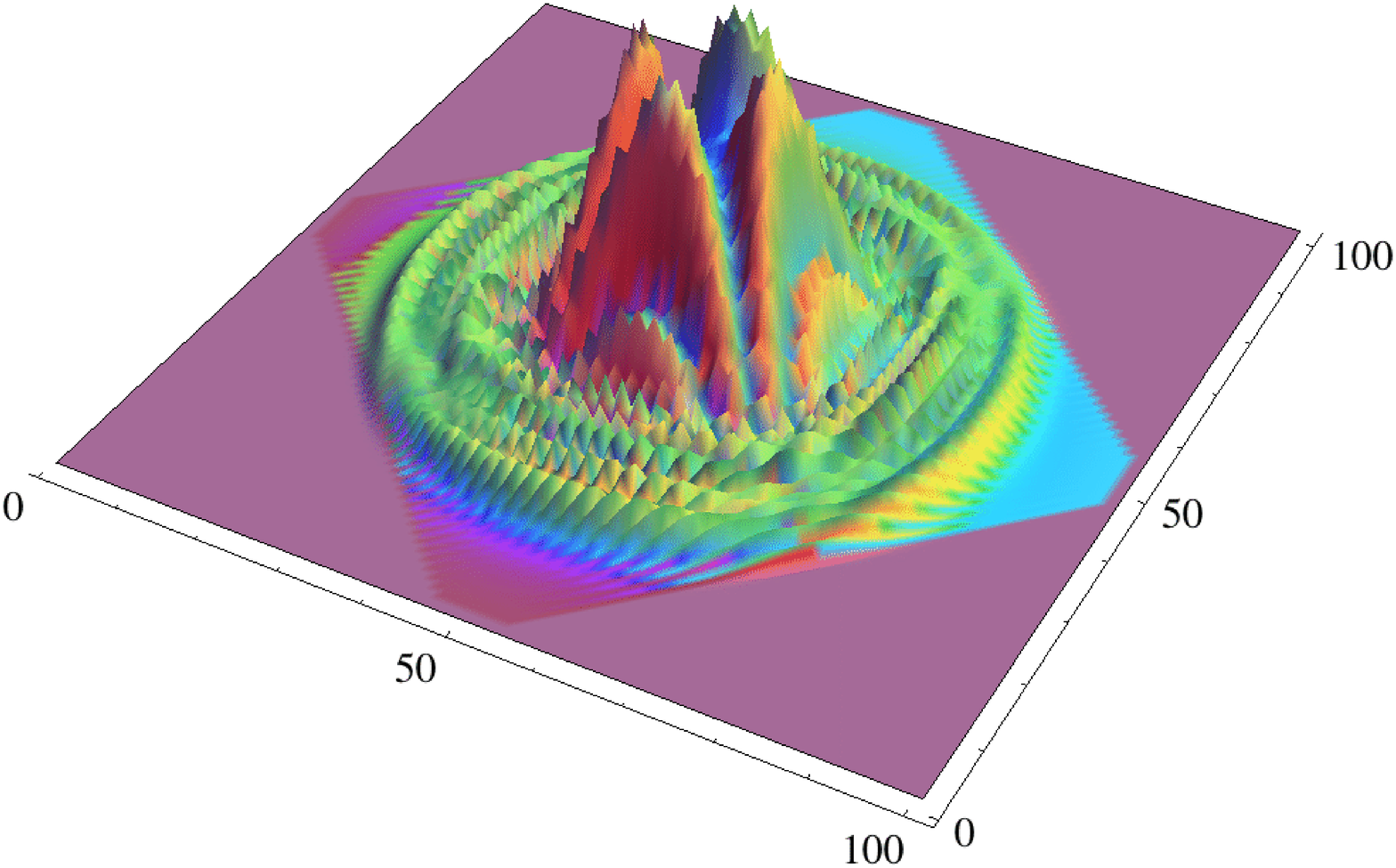}\\
(a)&(b)\\
\includegraphics[width=.46\textwidth]{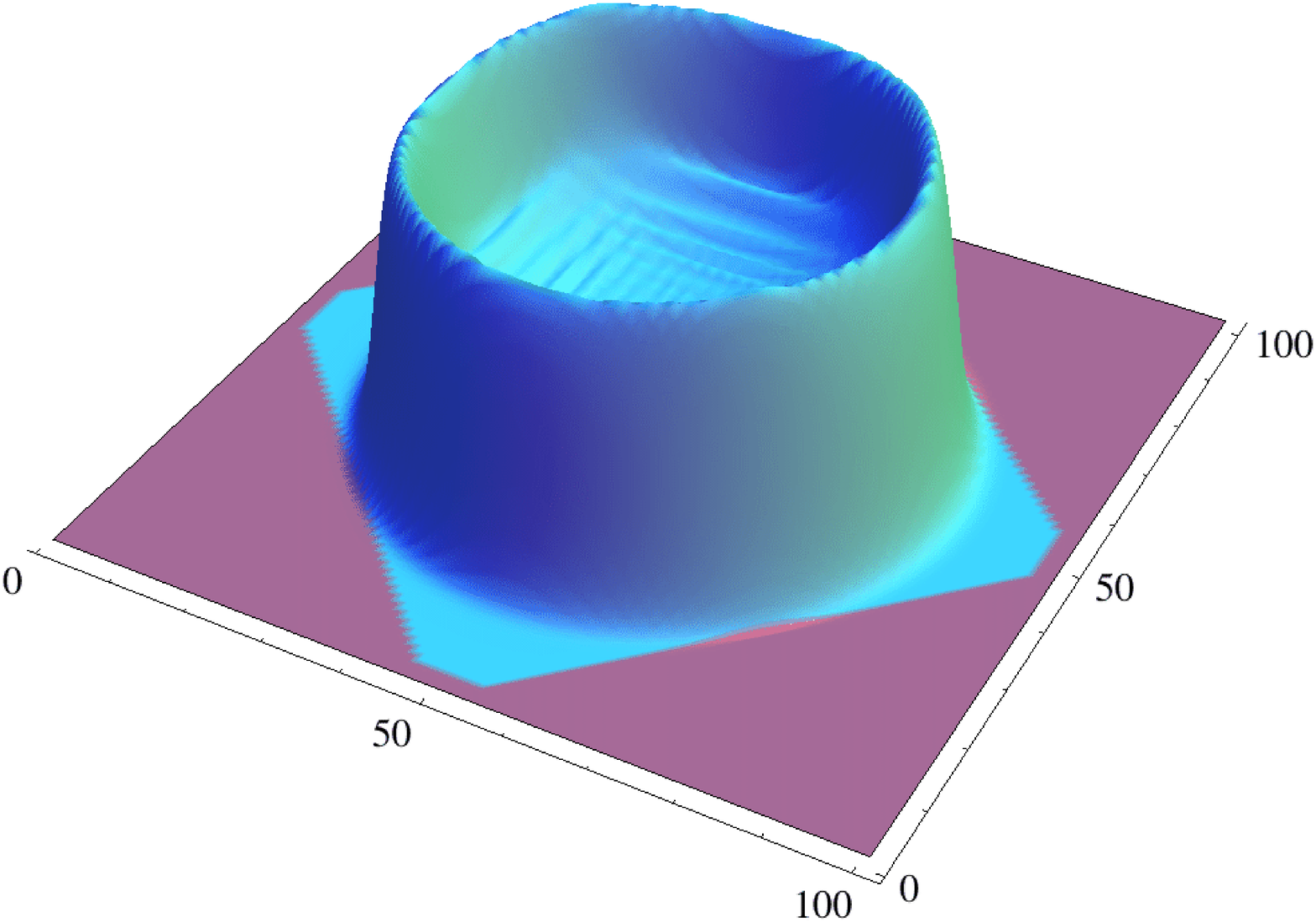}&\includegraphics[width=.46\textwidth]{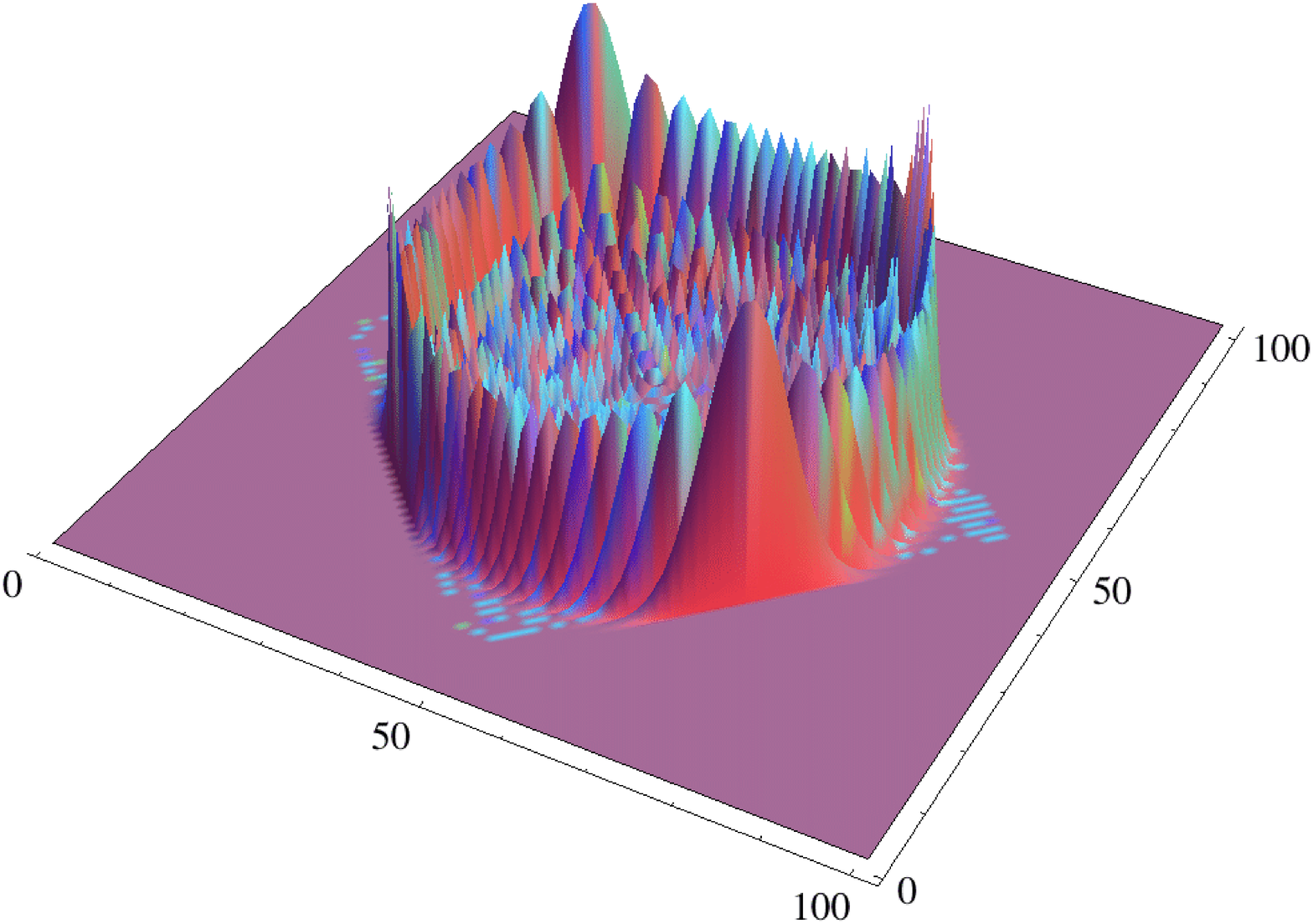}\\
(c)&(d)\\
\end{tabular}
\caption{How particles emerge from a Dirac automaton for $d=2$ space dimensions  \cite{MP}.  The
  height of the plot represents the probability of finding a particle with spin-up after 45 Planck
  times, for different initial states. Colors represents the spin state. (a,b,c): Gaussian
  wavepacket with $\Delta_x=\Delta_y=2l_P$, (d) state localized over a single Planck cell. Fig (a)
  high relativistic momentum, Fig. (b) ultrarelativistic momentum at Brillouin zone border, Figs.
  (c,d) zero momentum.  (Reprinted from Ref.  \cite{myFQXi12}).}\label{f:QCA2d}
\end{figure}

\subsection{Emergence of classical mechanics}
The Hamiltonian for the classical field theory corresponding to the quantum automaton can be derived
from the unitary operator of the automaton  \cite{Vaxjo2010,pla,myFQXi11}, and this is a
``classicalization rule'' of the kind already mentioned. Customary quantum particles are the mentioned
Schwartz-class smoothly-varying coherent superpositions of localized states of a single cell with
constant relative phase $\exp(i k_\alpha)$ between neighboring automaton cells in the $\alpha$
direction, $\vec{k}$ corresponding to the particle momentum. The classical trajectory is the
``typical path'' along the quantum network, i.e the path with maximum probability of the Gaussian
packet.

\subsection{Energy and momenta are bounded in the digital world} 
The maximum momentum is simply given by the De Broglie relation $\hbar\pi/l_P$. For Fermions, we can
have only one particle and one antiparticle per Planck cell, and the bound on how much energy can be
crammed into a unit of space is determined by the maximum energy per particle, which cannot be more
that $\hbar\pi t_P^{-1}=6.14663*10^9$J (a huge energy!).  This is the energy for achieving 2 ops
 \cite{Lloyd} of the automaton during the Planck time, as given by the Margulus-Levitin theorem
 \cite{Margolus-Levitin} (each step of the automaton is obtained with two rows of quantum gates).

\subsection{Spin-statistics}
The quantum automata for $d=1$ and $d=2$ can describe any statistics: the two particle states can be
both symmetrized or antisymmetrized, or worked as anyons. The derivation of the automaton for
dimension $d=3$ will be crucial for establishing the relation with spin, through covariance of the
unitary matrix under discrete rotations of the interaction lattice. For dimension $d=3$ there seems
to exist two inequivalent automata describing the flow of information \cite{MP}: hopefully they will
be related to the two Fermi and Bose statistics, respectively. In such way we will have an automaton
also for the Klein-Gordon massive field.

\section{A quantum-digital space-time}
The quantum nature of the automaton is crucial for the emergence of space-time. Indeed, there are
two crucial motivations against using a classical automaton.

The first point against a classical automaton is that with a classical automaton one cannot have
isotropic space emerging from an homogeneous classical causal network, due to the {\em Weyl Tile
  argument}  \cite{Weyl}: we count the same number of tiles in a square lattice both along the
diagonal and in the direction of the square sides. Thus Weyl asks: where the $\sqrt{2}$ comes from?
Indeed, the maximal speed of information in bcc-lattice automaton, as in the Dirac case, would be
faster by a factor $\sqrt{2}$ or $\sqrt{3}$ along diagonals than along lattice axes, ending up with
an anisotropic space for any homogeneous lattice  \cite{Fritz}. The problem is clearly not cured by
the continuum limit.  Instead, in a quantum network isotropy is recovered through quantum
superpositions of different paths (see e.g. Fig. \ref{f:QCA2d}c), and we have again isotropy of
max-speed in the relativistic regime of small momenta (Fig. \ref{f:gv}), whereas anisotropy would be
in principle visible only in the ultrarelativistic regime of huge momenta (Figs.
\ref{f:gv},\ref{f:QCA2d}b) or for ultra-localized states (Fig. \ref{f:QCA2d}d). In a similar manner
the quantum nature of the network provides the mechanism for restoration of all continuum symmetries
in the relativistic regime.  The digital version of Lorentz transformations for a classical
homogeneous causal network can be found in Ref.   \cite{tosini}: the usual Lorentz covariance cannot
be restored from them. Recovering Lorentz covariance from a classical causal network (i.e.
describing a causal ordering partial relation) conflicts with the homogeneity principle, and needs a
random topology, as in the causal-set program of Sorkin  \cite{sorkin}.

The second point against a classical automaton is that quantum superposition of localized states
provides a mechanism for directing information in space, by having constant relative phase between
neighboring systems in the superposition, thus giving momentum to the information flow. Such
mechanism is not possible in a classical causal network. It is the interplay between quantum
coherence and nonlocality that plays the crucial role of keeping information going along a desired
direction with minimal spreading, a task that can be accomplished inefficiently by a classical
automaton.

\section*{Conclusions and perspectives} 
The most important question is now where gravity comes from in the QCA: the answer would definitely
assess the power of the QCA framework, which, as mentioned in the introduction, is very natural and
promising for quantum gravity, and for a long list of reasons. I still don't know the answer, but I
believe that the equivalence principle must be rooted in the automaton mechanism itself, namely the
gravitational force must emerge at the level of the Dirac free theory, since it defines the inertial
mass. Besides, we have seen how the gravitational constant is related to the universal constants of
the automaton, and such relations is very enticing. Whereas this doesn't happen to be the case for
customary quantum field theory, it can possibly happen in the quantum automaton theory, as a tiny
effect from the Planckian dynamics, emerging in the ``thermodynamic'' limit as a purely {\em
 quantum-digital effect}.  Moreover, as already mentioned, the digital nature of the quantum
automaton makes it the natural scenario for the generalized holographic principle at the
basis of the Jacobson-Verlinde idea of gravity as entropic force  \cite{jacobson,verlinde}. The
hypothesis of gravity as a quantum-digital effect is very fascinating: it would mean we are indeed
experiencing the quantum-digital nature of the world in every moment and everywhere: through
gravity!

\end{document}